# Retention in STEM: Factors Influencing Student Persistence and Employment

**Linli Zhou, Ph.D.**
Institutional Research Analyst
Lasell University
lzhou@lasell.edu

**Damji Heo Stratton, Ph.D.**
Data Analyst
University of Missouri System
dhsdfn@umsystem.edu

**Xin Li, Ph.D.**
Postdoctoral Researcher
New York University, Shanghai
xl4749@nyu.edu

**Abstract:**
**This study utilizes data from the Baccalaureate and Beyond Longitudinal Study (B&B: 16/17) to explore factors associated with the likelihood of students' employment in STEM fields one year after graduation. We examined various factors related to students' individual characteristics (e.g., gender, race, and financial situation), institutional experiences (e.g., major, academic standing, research involvement, internships, extracurricular activities, and undergraduate practicum), and institutional and national trends. The results indicate lower STEM employment likelihood for minority groups and students with academic probation. The findings also highlight the positive impact of undergraduate practicum and job relevance to major on STEM employment likelihood. On the contrary, career services were negatively associated with the likelihood of students' STEM occupation choice, suggesting potential shortcomings in STEM job preparation within these services. The study provides valuable insights and actionable recommendations for policymakers and educators seeking to increase diversity and inclusion in STEM fields, suggesting the need for more efficient and tailored educational interventions and curriculum development.**

## Introduction

The science, technology, engineering, and mathematics (STEM) workforce drives innovation and productivity in the global economy, and the demand for STEM workers continues to grow (U.S. Bureau of Labor Statistics, 2021). Higher education scholars have focused on the STEM Education retention pipeline, which refers to the process of attracting, preparing, and retaining students in STEM fields from elementary school to postsecondary education (Roemer, 2020). However, the leaky pipeline is still ongoing from STEM college graduates to the STEM workforce, and factors that influencing these phenomena have not been adequately explored (Zhou et al, 2023).

In order for higher educational institutions to provide their students with adequate support, it is crucial to find factors associated with STEM college graduate's retention in the STEM workforce and how they differ by gender and race. For instance, policymakers and other stakeholders at higher educational institution could conduct more efficient and tailored educational intervention and curriculum development. Moreover, understanding national trends can help stakeholders assess where individual institutions stand and more accurately gauge the quality of their educational services and support. Thus, we examined factors related to STEM college graduates' choices and transition to the workforce.

In this study, we focus on the following research question: How do students' characteristics, college experiences, and institutional and social support associate with their choice of a STEM occupation?

Employment in a chosen field is a critical outcome that students aim for as they progress through their undergraduate programs, laying the foundation for their future. To gain a deeper understanding of this process, researchers have identified various factors that influence students' employment prospects. These factors encompass personal background, self-efficacy, and contextual elements like educational experiences (Beier et al., 2019; Lent, Brown, & Hackett, 2002). Consequently, as the theoretical framework for our study, we adopted the Social Cognitive Theory and Social Cognitive Career Theory (SCCT) framework (see Figure 1, adapted from Bandura, 1977). SCCT has been widely employed to elucidate students' selection of majors and career paths (Kanny et al., 2014). It is a well-established theory that investigates the interplay between individuals, their contexts, and the social cognitive factors involved in career development (Li et al., 2019). Within SCCT, personal, contextual, and experiential factors exert influence on three fundamental "building blocks" of individual career decisions. Firstly, demographic and academic characteristics can shape one's career interests and choices. Additionally, engaging in career-related activities can enhance one's interest and confidence in their career path. Conversely, without such involvement in career-related practices and experiences, interest development and decision-making may be limited. Building upon the SCCT framework, we examined various factors pertaining to students' individual characteristics (gender, race, financial situation), institutional experiences (undergraduate major, academic standing, research involvement, internships, extracurricular activities, and practicum), and institutional and social support (employment services, as well as support from family and friends).

We conducted logistic regression analyses to examine how these factors related to the likelihood of college graduates being employed in a STEM field. We used data from the Baccalaureate and Beyond Longitudinal Study (B&B:16/17), which contains variables of interest for this project, including students' characteristics, undergraduate experiences, and institutional and social support.

We found that students from minority groups were 20% less likely to choose STEM careers than White students. Students who had ever been on academic probation were found to be 29% less likely to be employed in a STEM field. The result infers the disparity in workforce access and resources for underrepresented minority groups, and the need to support academic success for all students, particularly those at risk of academic probation. Meanwhile, we found students with a STEM degree were 3.83 times more likely to be employed in the STEM workforce. Students who participated in undergraduate practicum were 1.35 times more likely to have a STEM occupation. Students who had jobs related to their major were 1.65 times more likely to be employed in the STEM workforce. These results highlight the importance of STEM education, particularly experiential learning and internship experiences related to one's major. Lastly, the results revealed that students whose institutions offer employment services were approximately half as likely to be employed in STEM fields, compared to those who attended institutions that do not offer employment services. Also, students who used career counseling services were 22% less likely to be employed in STEM fields. The finding suggests that career counseling services may not focus on STEM job preparation, or the service is not enough to prepare students for employment in STEM fields.

This study has implications for diversity and inclusion in STEM education. It reminds educators to support minority groups and students with academic difficulty to ensure their access to STEM majors. The project also encourages higher education leadership to implement experiential program design, and to focus on internship and practicum experiences related to students' majors. We also suggest that career services in higher education institutions focus on STEM career preparation and explore necessary resources beyond the current counseling services to improve the placement of college graduates in the STEM workforce.

**Method**

**Data and Sample**

The project uses data from the Baccalaureate and Beyond Longitudinal Study (B&B:16/17), a national survey that tracked outcomes of 2015–16 bachelor's degree recipients one year after graduation. The survey focused on college outcomes, including undergraduate enrollment experiences, post-baccalaureate enrollment, characteristics of the first postbaccalaureate job, financial well-being, and student loan debt and repayment (Thomsen, et.al., 2020).

The B&B:16/17 contains a nationally representative sample. The survey respondents were identified from the 2015–16 National Postsecondary Student Aid Study (NPSAS:16), which sampled approximately 122,000 students from 2,000 US institutions. The B&B:16/17 surveyed 26,500 qualified students who finished their bachelor's degree requirements between July 1, 2015, and June 30, 2016, and received a bachelor's degree by June 30, 2017. Of the qualified students, about 20,000 responded to the B&B:16/17 survey, representing a 71% weighted response rate.

In the current study, we focused on the survey questions about students' post-baccalaureate jobs and career choices. We examined the occupation types of 2015–16 bachelor's degree recipients' first full-time jobs within one year of bachelor's degree completion. Among all the survey respondents, we selected students who held a college degree in science, technology, engineering, and mathematics (STEM) fields (Chen, 2009). Thus, this study examined a data sample of 4,000 STEM college graduates.

**Analysis Tool**

During our data analysis, we used PowerStats, a web-based software application officially recommended by NCES for analyzing the B&B and other surveys they conducted. PowerStats ensures estimates represent the target population by generating weights for survey samples (Radwin, et.al., 2018). These weights account for nonresponse, subsampling of potential B&B:16/17-eligible students, multiplicity at the student level and unknown student eligibility for NPSAS:16, and unequal probability of selection of institutions and students in the NPSAS:16 sample (Velez, et.al., 2019). PowerStats also generates design-adjusted standard errors for statistical significance testing using replicate weights produced with balanced repeated replication, jackknifing, or bootstrapping (Velez, et.al., 2019).

**Regression Model**

This study uses logistic regression to estimate the probability of students' employment in STEM fields using the selected exploratory variables. Accordingly, the following model was used:

$Y(B1STEMOC) = \alpha + \beta_1 B1GENDER + \beta_2 RACE + \beta_3 OWEAMT1 + \beta_4 B1MAJORS4Y + \beta_5 B1MAJCHO + \beta_6 B1EXPAP + \beta_7 B1UGRESEARCH + \beta_8 B1PRACT + \beta_9 B1CLB + \beta_{10} STUSERV1 + \beta_{11} B1UGCARSRVS2 + \beta_{12} B1LKFAM + \beta_{13} B1LKINT + \beta_{14} JOBMAJOR3 + \varepsilon$

We selected the exploratory variables based on previous literature and the theoretical framework discussed in the previous sections. To make the model concise, we selected independent variables (see Appendix A) based on a correlation matrix (see Appendix B to Appendix F). We identified and excluded correlated variables based on a correlation threshold of 0.17. Those correlated variables possibly measure similar aspects. When two variables are correlated, the variable with a stronger correlation with the dependent variable is selected.

The dependent variable in the model is 'Ever employed in a STEM occupation' (B1STEMOC), a categorical variable indicating whether the respondent has ever worked in a STEM field within 12 months after completing a bachelor's degree. The independent variables include personal and institutional experiential variables. A list of variables and their percentages of distribution are displayed in Table 1.

The model contains several independent variables, including students' personal characteristics and undergraduate experiences. The first set of independent variables that the model included are students' characteristics, such as gender (B1GENDER), a categorical variable with values of three levels: male (42%, reference group), female (57%), and LGBTQ+ (1%) (transgender, queer, unsure, or more than one gender). We included race (RACE), a categorical variable that has White (65%, reference group) and minority groups (35%) (Black or African American, Hispanic or Latino, Asian, American Indian or Alaska Native, Native Hawaiian/other Pacific Islander, and More than one race). In addition, the model included owed undergraduate loans (OWEAMT1) that measure students' financial situation with two categories: above loan (26%) and below average loan (74%, reference group).

The second set of independent variables measures students' undergraduate experiences. Undergraduate major (B1MAJORS4) has two categories: STEM (including computer and information sciences, engineering and engineering technology, and biological and physical science, science technology, math, and agriculture) and non-STEM (including general studies and other, social sciences, humanities, health care fields, business, and education). Satisfaction with undergraduate majors (B1MAJCHO) has two categories: satisfied (very satisfied and satisfied) and dissatisfied (very dissatisfied and dissatisfied, reference group). Other undergraduate experiences variables include: "ever on academic probation" (B1EXPAP), "participated in undergraduate research" (B1UGRESEARCH), "undergraduate practicum" (B1PRACT), "involved in extracurricular groups" (B1CLB), "completed an internship" (B1LKINT), and "job ever related to major" (JOBMAJOR3), which are all binary variables ("No" is the reference group) that measures student institutional experiences.

Lastly, we included career-related resources and support, such as 'utilized career counseling' (B1UGCARSRVS2), 'institutions offering employment services' (STUSERV1), and 'talked to friends or family members' (B1LKFAM). These are binary variables measuring institutional and social support for students.

**Results**

Table 2 shows the results of descriptive statistics about the group differences for STEM and Non-STEM occupations among different racial, gender, financial, academic performance, and experiences groups. Besides the descriptive statistics, this section mainly focuses on the results from the logistic regression, including the coefficients, odds ratios, and significance levels for each predictor in the model. Table 3 displays details about the coefficients, Odds Ratio, and *p*-value of each independent variable.

Overall, the model has a negative log-likelihood (Pseudo R2) of 0.076, suggesting that the logistic regression model fits the data very well (Minitab 20 Support, 2021; Sobel & Shapiro, 1982; Tovar, 2020; Towers, 2019). The likelihood ratio (Cox-Snell) maximum is 0.6852, meaning the model explains about 68.52% of the variation in the dependent variable and has a moderate predictive power in explaining the outcome variable. Other model fit measures can be found in Table 4.

**Demographic Characteristics and Financial Characteristics**

We found that although there was no significant difference in the probability of STEM employment for female ($p = 0.31$) and LGBTQ+ ($p = 0.76$) students compared to male students, there was a significant difference for minority groups, who were 20% less likely to choose STEM careers than White students ($p = 0.02$, Odds Ratio = 0.8). We did not find a significant difference made by a student's financial situation for their STEM career choice ($p = 0.32$).

**Undergraduate Experiences**

Our results reveal that having a STEM undergraduate degree increases a student's likelihood of being employed in the STEM workforce by 3.83 times ($p < 0.01$). However, no significant association was found between students' satisfaction with their majors and their STEM employment ($p = 0.57$). Moreover, students with academic probation experiences were found to be 29% less likely to be employed in a STEM field ($p = 0.04$, Odds Ratio = 0.71). Furthermore, while students who participated in an undergraduate practicum were 1.35 times more likely to have a STEM occupation ($p = 0.01$, Odds Ratio = 1.35), extracurricular experiences did not significantly raise the likelihood of STEM employment ($p = 0.72$, Odds Ratio = 1.04). Students' research experiences did not significantly improve their probability of STEM employment ($p = 0.06$, Odds Ratio = 1.24) either. We also found that students who had jobs related to their major were 1.65 times more likely to be employed in the STEM workforce ($p < 0.01$, Odds Ratio = 1.65). However, the completion of an internship had no significant relationship with students' STEM employment ($p = 0.43$). The overall results suggest the importance of relevant work experiences in preparing the STEM workforce.

**Institutional Services and Social Support**

The results indicate that students whose institutions offer employment services were approximately half as likely to be employed in STEM fields, compared to those who attended institutions that do not offer employment services ($p < 0.01$, Odds Ratio = 0.46). Meanwhile, we found students who used career counseling services were 22% less likely to be employed in STEM fields ($p = 0.04$). In addition to institutional career services, we examined students' social support and found that talking to friends and family did not significantly contribute to their likelihood of STEM employment ($p = 0.93$, Odds Ratio = 0.99).

**Discussion**

This study significantly contributes to the existing literature by providing a comprehensive analysis of various factors influencing students' likelihood of entering the STEM workforce. It offers both confirmation of and new insights into these factors, which will aid policymakers, educators, and career counselors in enhancing career preparation strategies for students in STEM fields.

We found that minority groups have a significantly lower likelihood of STEM employment compared to the white student group. This aligns with Landivar's (2013) findings, which reveal disparities in STEM employment opportunities for different racial groups. Meanwhile, students on academic probation were less likely to enter the STEM workforce. Similar results were previously reported by Tovar & Simon (2006), which suggested the importance of academic support, while our study extended the importance of support for academic at-risk students for career preparation.

We found that participating in undergraduate practicum can better prepare students to enter the STEM workforce. This finding aligns with Veenstra's (2014) emphasis on experiential learning for STEM education. The result indicating that having a job related to one's major could enhance the likelihood of STEM career employment generally aligns with research on how internship and practical experiences can prepare students for the STEM workforce (e.g., Veenstra, 2014). However, we also found that research experiences, general internships, and extracurricular activities do not significantly improve students' likelihood of STEM employment. This finding indicates that the benefit of research, internships (unrelated to one's major), and extracurricular activities may work for other fields (e.g., Strapp & Farr, 2010), but they may not help prepare students seeking employment in STEM fields.

On the other hand, we found that career services are negatively associated with students' likelihood of having a STEM occupation, which deviates from previous literature emphasizing the importance of career counseling for student career preparation (e.g., Schmidt & Rokutani 2012). One possible explanation is that institutions offering employment services may not focus on STEM job preparation, or students who utilize career counseling services may not be focusing on STEM job preparation. It is also possible that institutions that do not offer employment services may have other resources that prepare students for STEM careers, and students who did not utilize career counseling may rely on other resources to help their STEM job hunting.

We also found that social networks do not significantly improve students' likelihood of preparing for STEM occupations, which is inconsistent with previous literature emphasizing students' social capital and networking in career preparation (e.g., Ceglie & Settlage, 2016). A plausible explanation is that talking to family and friends may not be as effective in helping students navigate the complex process of finding a STEM career.

**Conclusion and Recommendations**

This project contributes to our understanding of the factors related to students' likelihood of being employed in STEM fields. Additionally, this study provides valuable insights for policymakers

---

and educators seeking to increase diversity and inclusion in the STEM workforce. Based on our findings, we suggest actionable recommendations and directions for potential future research studies as follows.

**Supporting Minority Students and Students with Academic Difficulty**

From the findings of our study, we recognize that being part of a minority group can limit one's access to the STEM workforce. We thus advocate for more efforts to represent diverse populations in STEM occupations, such as providing equal career preparation resources for underrepresented groups. Also, this study revealed that academic probation could result in a lower likelihood of having a STEM occupation. We therefore encourage educators to prioritize efforts to support academic success for all students, particularly those at risk of academic probation.

**Providing Practicum and Internships Related to Student Major**

This study reveals a positive correlation between practicum experiences and the likelihood of entering the STEM workforce. Therefore, we encourage undergraduate program designers to incorporate more experiential learning opportunities in STEM education to promote the retention of college graduates in the STEM workforce. In addition, we found that having internships related to one's major could improve the likelihood of obtaining a STEM occupation. We recommend undergraduate institutions provide more opportunities for students to gain work experience related to their major to promote STEM workforce preparation for college students.

**Redesigning Institutional Career Services**

Our results reveal that both employment services at higher education institutions and student utilization of career services do not improve their probability of being hired in STEM fields. The results suggest that institutional career services may not adequately prepare students interested in STEM fields. We thus urge institutions that currently offer employment services to reconsider their career services setup and improve the design of these services based on students' needs. We recommend career counselors collect more feedback from students interested in STEM careers and STEM employers about how they can better support them regarding STEM workforce preparation.

**Directions for Potential Further Research**

Our study provides valuable insights into the factors related to students' likelihood of being employed in STEM fields. However, there are several limitations. First, our data is limited to self-reported responses one year after bachelor completion. We cannot establish causality or account for changes in student experiences or attitudes over time. Second, we rely on binary categorization of variables, such as whether a student was involved in a particular experience. This method does not capture the full complexity and diversity of student experiences that could influence their STEM career choices. Third, the PowerStats analytical tool does not allow us to perform a collinearity analysis to check for multicollinearity among our predictor variables. Instead, we relied on correlation analysis to select variables for inclusion in our logistic regression model. While this approach is common in many studies, it may have led to the exclusion of influential variables or the inclusion of redundant variables, potentially affecting the accuracy of our model. Meanwhile, PowerStats cannot perform a factor analysis or include interaction terms in our regression model. This limitation prevented us from identifying underlying latent factors that may have influenced our results.

The limitations mentioned above suggest the need for future research to expand upon our findings. For example, future studies may explore potential causal relationships between the factors we examined and students' employment in STEM fields over a longer period. Also, future studies could explore more comprehensive measures to capture the complexity of student experiences. In addition, further investigations into the specific types of practicum opportunities and work experience that are most helpful for preparing for a STEM career could be conducted. Lastly, future studies can include examining potential interventions to address challenges, such as in career services utilization, to improve students' experiences and workforce outcomes in STEM education.

---

**Acknowledgments**

We thank Ashley Gerhardson, Grace Shin, Johnny Woods Jr, and Tien Ling Hu for their inspiration on this topic during the NCES Data Institute. We appreciate the learning opportunity about national surveys from participating in the NCES Data Institute. We also thank Steven Sherrin and Eric Lanthier for their reviews of the paper and discussions around college student retention and graduation outcomes.

## Tables

**Table 1**: *Percentage Distribution of Independent and Dependent Variables*

|  | Yes | No | Weighted Sample Size |
|---|---|---|---|
| STEM Occupation | 29% | 71% | 1,839,449 |
| Race: Minority group | 35% | 65% | 2,038,555 |
| Owed undergrad loan above average | 26% | 74% | 1,839,448 |
| Undergraduate Major: STEM | 21% | 79% | 1,839,448 |
| Satisfied with undergraduate major | 94% | 6% | 1,740,820 |
| Ever on academic probation | 8% | 92% | 1,839,448 |
| Participated in undergraduate research | 26% | 74% | 1,839,449 |
| Participated in undergraduate practicum | 15% | 85% | 1,839,449 |
| Involved in extracurricular group | 57% | 43% | 1,839,447 |
| Utilized career counseling | 22% | 78% | 1,839,449 |
| Talked to friends or family members | 63% | 37% | 1,238,126 |
| Completed an internship | 10% | 90% | 1,238,126 |
| Job major ever related to major | 53% | 47% | 1,557,575 |
| Institution offers employment services | 96% | 4% | 1,833,162 |

**Table 2**: *Student Characteristics by STEM Occupation*

| Student Characteristics | STEM Occupation | Non-STEM Occupation |
|---|---|---|
| Total | 29% | 71% |
| Gender: Male | 33% | 67% |
| Gender: Female | 25% | 75% |
| Gender: LGBTQ+ | 26% | 74% |
| Race: White | 30% | 70% |
| Race: Minority | 26% | 74% |
| Owed undergrad loan: Below average | 29% | 71% |
| Owed undergrad loan: Above average | 27% | 73% |
| Undergraduate Major: STEM | 53% | 47% |
| Undergraduate Major: non-STEM | 22% | 78% |
| Satisfied with undergraduate major | 29% | 71% |
| Dissatisfied with undergraduate major | 24% | 76% |
| Not ever on academic probation | 29% | 71% |
| Ever on academic probation | 21% | 79% |
| Participated in undergraduate research | 34% | 66% |
| Did not participate in undergraduate research | 27% | 73% |
| Participated in undergraduate practicum | 28% | 72% |
| Did not participate in undergraduate practicum | 29% | 71% |
| Involved in extracurricular group | 29% | 71% |
| Did not involve in extracurricular group | 28% | 72% |
| Institution offers employment services | 28% | 72% |
| Institution does not offer employment services | 42% | 58% |
| Did not utilize career counseling | 29% | 71% |
| Utilized career counseling | 27% | 73% |
| Did not talk to friends or family members | 29% | 71% |
| Talked to friends or family members | 26% | 74% |
| Completed an internship | 24% | 76% |
| Did not complete an internship | 27% | 73% |
| Job ever related to major | 36% | 64% |
| Job not ever related to major | 22% | 78% |

**Table 3**: *Coefficients and Odds Ratio of Different Independent Variables Groups Compared to the Reference Group for the Probability of Ever Employed in STEM Occupations*

*214*| Logistic Regression Results | Odds Ratio | b | b Standard Error | t | p-value |
|---|---|---|---|---|---|
| Intercept | 0.42 | -0.86 | 0.26 | -3.37 | 0.00 |
| Race (reference: White) | | | | | |
|   Minority | 0.80 | -0.23 | 0.09 | -2.40 | 0.02 |
| Gender (Reference: Male) | | | | | |
|   Female | 0.91 | -0.09 | 0.09 | -1.03 | 0.31 |
|   LGBTQ+ | 0.89 | -0.12 | 0.39 | -0.30 | 0.76 |
| Owed undergrad loan: Above average | 1.09 | 0.09 | 0.09 | 0.99 | 0.32 |
| Undergraduate Major: STEM | 3.83 | 1.34 | 0.10 | 12.98 | 0.00 |
| Satisfied with undergraduate major | 1.09 | 0.09 | 0.16 | 0.56 | 0.57 |
| Ever on academic probation | 0.71 | -0.34 | 0.17 | -2.08 | 0.04 |
| Participated in undergraduate research | 1.24 | 0.21 | 0.11 | 1.90 | 0.06 |
| Participated in undergraduate practicum | 1.35 | 0.30 | 0.12 | 2.61 | 0.01 |
| Involved in extracurricular group | 1.04 | 0.04 | 0.10 | 0.36 | 0.72 |
| Institution offers employment services | 0.46 | -0.78 | 0.20 | -3.91 | 0.00 |
| Utilized career counseling | 0.78 | -0.25 | 0.12 | -2.09 | 0.04 |
| Talked to friends or family members | 0.99 | -0.01 | 0.09 | -0.09 | 0.93 |
| Job ever related to major | 1.65 | 0.50 | 0.10 | 5.15 | 0.00 |
| Completed an internship | 0.89 | -0.12 | 0.15 | -0.80 | 0.43 |

**Table 4**: *Logistic Regression Model Fit Measures*

| | |
|---|---|
| Negative log-likelihood (Pseudo R2) | 0.076 |
| Log-likelihood (intercept only) | -353,443.64 |
| Log likelihood (full model) | -326,569.92 |
| Likelihood Ratio (Cox-Snell) | 0.0841 |
| Likelihood Ratio (Cox-Snell) Maximum | 0.6852 |
| Likelihood Ratio (Estrella) | 0.0873 |
| Degrees of Freedom | 200 |
| Number of Categories | 15 |

*Zhou, L., Stratton, D. H., & Li, X. (2023). Retention in STEM: Factors Influencing Student Persistence and Employment. The Proceeding of the 19th Annual National Symposium on Student Retention, 205–217.*

---

**Figures**

**Figure 1**: *Social Cognitive Career Theory (SCCT) Model (Lent et al., 1994; Kanny et al., 2014.)*

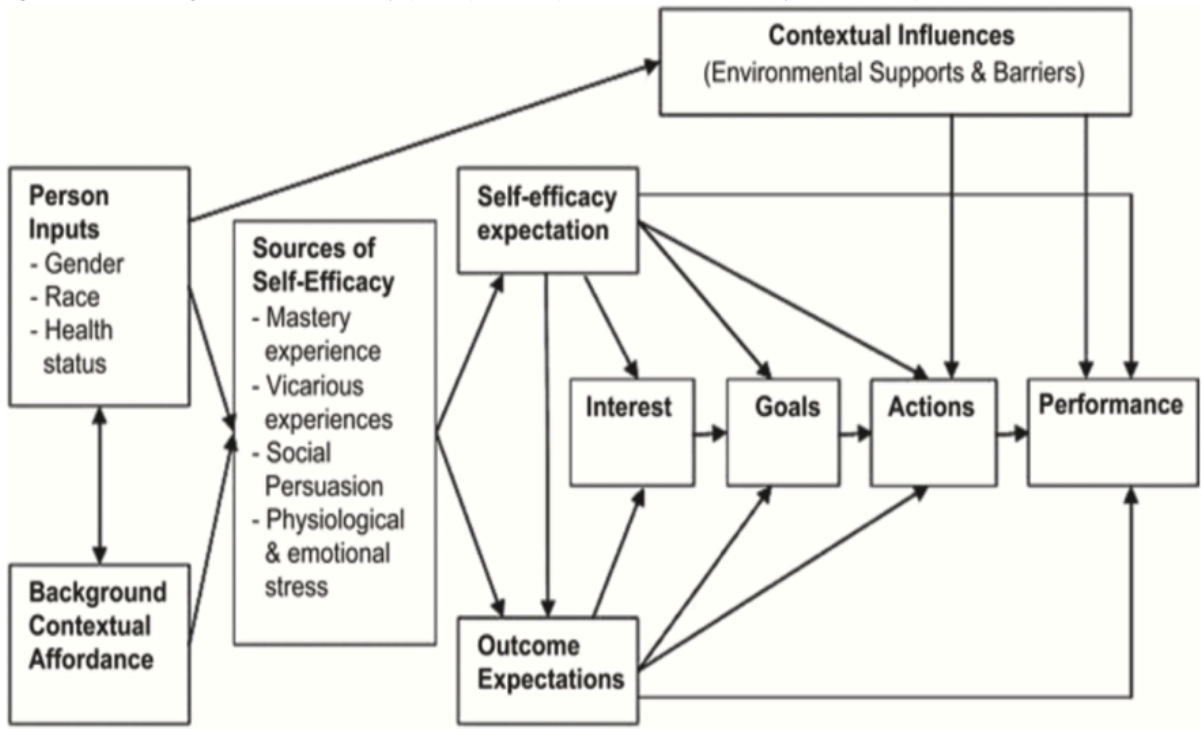

---

# Appendix

**Appendix A**: *Model Independent Variables Selection based on Collinearity Analysis*

| Selected Variables for Regression Model | Excluded Variable (Correlation Analysis) |
|---|---|
| Dependent Variable:<br>- Ever employed in a STEM occupation (B1STEMOC)<br>Independent Variables:<br>1. Gender (B1GENDER)<br>2. Race (RACE)<br>3. Amount still owed on undergraduate loans (OWEAMT1)<br>4. Field of study: Undergraduate (B1MAJORS4Y)<br>5. Satisfaction with undergraduate major (B1MAJCHO)<br>6. Ever on academic probation (B1EXPAP)<br>7. Participated in undergraduate research (B1UGRESEARCH)<br>8. Participated in an undergraduate practicum (B1PRACT)<br>9. Involved in extracurricular club or group (B1CLB)<br>10. Institution offers employment services (STUSERV1)<br>11. Utilized career service counseling (B1UGCARSRVS2)<br>12. Talked to friends or family members (B1LKFAM)<br>13. Completed an internship (B1LKINT)<br>14. Job ever related to major (JOBMAJOR3) | Correlated with OWEAMT1:<br>- Importance of job wage<br>Correlated with B1UGRESEARCH:<br>- Ever on academic probation<br>Correlated with B1PRACT:<br>- Participated in a co-operative experience<br>Correlated with B1CLB<br>- Held a formal leadership role<br>- Participated in culminating senior experience<br>Correlated with STUSERV1:<br>- NPSAS institution control<br>- Selectivity<br>- Institution offers placement services<br>Correlated with B1UGCARSRVS2<br>- Participated in mock interviews<br>- Participated in career/job fairs<br>- Utilized alumni network<br>- Utilized career service assessments<br>- Utilized resume or cover letter assistance<br>- Utilized career service job database<br>Correlated with B1LKFAM<br>- Talked to coworkers or mentors<br>- Talked to faculty members or alumni<br>- Talked to friends or family members |

**Appendix B**: *Correlation Variables for Participated in an Undergraduate Practicum (B1PRACT) and Involved in Extracurricular Club or Group (B1CLB)*

|  | B1PRACT | Co-operative experience |
|---|---|---|
| B1PRACT | 1.000 |  |
| Co-operative experience | 0.174 | 1.000 |

**Appendix C:** *Correlation Variables for Involved in Extracurricular Club or Group (B1CLB)*

|  | B1CLB | Held leadership role | Culminating senior experience |
|---|---|---|---|
| B1CLB | 1.000 |  |  |
| Held leadership role | 0.468 | 1.000 |  |
| Culminating senior experience | 0.104 | 0.117 | 1.000 |

**Appendix D:** *Correlation Variables for Institution Offers Employment Services (STUSERV1)*

|  | STUSERV1 | Public | Private nonprofit | Private for-profit | Institution offers placement services | Selective |
|---|---|---|---|---|---|---|
| STUSERV1s | 1.000 |  |  |  |  |  |
| Public institution | 0.207 | 1.000 |  |  |  |  |
| Private nonprofit institution | 0.017 | -0.877 | 1.000 |  |  |  |
| Private for-profit institution | -0.457 | -0.331 | -0.163 | 1.000 |  |  |
| Institution offers placement services | 0.495 | 0.204 | -0.067 | -0.288 | 1.000 |  |

--------------------------------------------------------------------------------

| | | | | | | |
|---|---|---|---|---|---|---|
| Selective | 0.4210 | 0.238 | 0.019 | -0.526 | 0.325 | 1.000 |

**Appendix E:** *Correlation Variables for Utilized Career Service Counseling (B1UGCARSRVS2)*

| | B1UGCARSRVS2 | Alumni network | Mock interview | Career fairs | Career assessment | Resume assistance | Job database |
|---|---|---|---|---|---|---|---|
| B1UGCARSRVS2 | 1.000 | | | | | | |
| Alumni network | 0.374 | 1.000 | | | | | |
| Mock interview | 0.367 | 0.287 | 1.000 | | | | |
| Career fairs | 0.496 | 0.381 | 0.425 | 1.000 | | | |
| Career assessment | 0.431 | 0.289 | 0.301 | 0.394 | 1.000 | | |
| Resume assistance | 0.568 | 0.380 | 0.470 | 0.645 | 0.397 | 1.000 | |
| Job database | 0.501 | 0.374 | 0.328 | 0.608 | 0.434 | 0.597 | 1.000 |

**Appendix F:** *Correlation Variables for Talked to Friends or Family Members (B1LKFAM)*

| | B1LKFAM | Talked to coworker/mentor | Talked to faculty/alumni |
|---|---|---|---|
| B1LKFAM | 1.000 | | |
| Talked to coworker/mentor | 0.344 | 1.000 | |
| Talked to c faculty/alumni | 0.240 | 0.297 | 1.000 |